\documentclass{emulateapj}
\usepackage{apjfonts}
\usepackage{graphicx}
\usepackage{psfig}
\usepackage{amssymb}
\newcommand{\begit}{\begin{itemize}}
\newcommand{\enit}{\end{itemize}}
\newcommand{\begen}{\begin{enumerate}}
\newcommand{\enen}{\end{enumerate}}

\newcommand       \be           {\begin{equation}}
\newcommand       \ee           {\end{equation}}
\newcommand       \bea          {\begin{eqnarray}}
\newcommand       \eea          {\end{eqnarray}}

\newcommand       \pc		{\,{\rm pc }}
\newcommand       \yr		{\,{\rm yr }}

\newcommand{\beqa}{\begin{eqnarray}} 
\newcommand{\eeqa}{\end{eqnarray}}

\shorttitle{SECULAR EVOLUTION OF BINARIES NEAR MASSIVE BLACK HOLES}
\shortauthors{Prodan et. al}
\begin{document}

\title{SECULAR EVOLUTION OF BINARIES NEAR MASSIVE BLACK HOLES: Formation of compact binaries, merger/collision products and G2-like objects}

\author{Snezana Prodan\altaffilmark{1}, Fabio Antonini\altaffilmark{1,2} \& Hagai B. Perets\altaffilmark{3}}

\altaffiltext{1}{Canadian Institute for Theoretical Astrophysics, 60
St.~George Street, University of Toronto, Toronto, ON M5S 3H8, Canada;
sprodan@cita.utoronto.ca, antonini@cita.utoronto.ca} 
\altaffiltext{2}{Departimento di Fisica, Universita' di Roma `La Sapienza', P.le A. Moro 5,
I-00185, Rome, Italy} 
\altaffiltext{3}{Deloro Fellow; Physics Department, Technion - Israel Institute of Technology, Haifa, Israel 32000}

\begin{abstract}
Here we discuss the
evolution of binaries around MBH in nuclear stellar clusters. We focus on their secular evolution
due to the perturbation by the MBH, while simplistically accounting
for their collisional evolution. Binaries with highly inclined orbits
in respect to their orbit around the MBH are strongly affected by
secular processes, which periodically change
their eccentricities and inclinations (e.g. Kozai--Lidov cycles). During
periapsis approach,
dissipative processes such as tidal friction may become highly efficient,
and may lead to shrinkage of a binary orbit and even to its merger.
Binaries in this environment can therefore significantly change their
orbital evolution due to the MBH third-body perturbative effects.
Such orbital evolution may impinge on their later stellar
evolution. Here we follow the secular dynamics of such binaries and
its coupling to tidal evolution, as well as
the stellar evolution of such binaries on longer time-scales. We find
that stellar binaries in the central parts of  NSCs are highly likely
to evolve into eccentric and/or short period binaries, and become
strongly interacting binaries either on the main sequence (at which
point they may even merge), or through their later binary stellar
evolution. The central parts of NSCs therefore catalyze the formation
and evolution of strongly interacting binaries, and lead to the enhanced
formation of blue stragglers, X-ray binaries, gravitational wave sources
and possible supernova progenitors. Induced mergers/collisions
may also lead to the formation of G2-like cloud-like objects such
as the one recently observed in the Galactic center.
\end{abstract}


\section{INTRODUCTION}

The majority of stars are believed to be in binaries or higher multiplicity systems, both in the field and in the dense stellar environments of globular clusters and galactic nuclei. In the inner parts of the
nuclear stellar cluster (NSC) of the Galactic center (GC; within $\sim1$
pc) the gravitational potential is dominated by the central massive
black hole (MBH). Binaries in the NSC are bound to the MBH and effectively
form hierarchical triple systems with the MBH (i.e. the binary orbit
around the MBH is the outer orbit of the triple).
If the orbit of a binary is highly inclined with respect to its orbit around the MBH, strong oscillations of the inner orbit eccentricity and mutual inclination are induced on secular timescale. The secular timescale is often shorter then the timescale over which gravitational interactions with background stars would significantly affect either the internal or the external orbit of the binary. These oscillations are known as Kozai--Lidov (KL) cycles \citep{1962P&SS....9..719L,1962AJ.....67..591K}. The induced high eccentricities could lead to strong interactions between the stellar binary components, producing significant orbital shrinkage, as well as potentially affecting the later binary stellar evolution. 

\citet[AP12]{2012ApJ...757...27A} explored the evolution of compact
binaries (white dwarfs, neutron stars and black holes) orbiting the MBH. 
AP21 followed the  coupled KL evolution and the gravitational
wave (GW) emission was followed, while considering the potentially limited lifetime
of binaries due their softening and final destruction through encounters
with stars in the nuclear cluster. It was shown that such coupled
evolution could significantly affect the binary evolution and enhance
the rate of GW sources formation, as well as change their characteristics,
in particular producing eccentric GW sources. Furthermore, AP12 discovered
the regime of quasi-KL evolution occurring for weakly hierarchical
triples. During such evolution, the eccentricity of the inner binary experiences significant changes on dynamical timescale due
to the perturbations from a third body.  Such changes in the eccentricity are no longer of purely secular nature and therefore one should be cautious when numerically treating such systems.  Some of the implications of this discovery are discussed in \citet{2014ApJ...781...45A} as well. 

Here we expand on AP12 and \citet{2010ApJ...713...90A} and explore the evolution of main-sequence
(MS) binaries near MBHs and environments similar to the NSC of the Milky-way. 
As we discuss in the following, the coupling of KL-cycles and tidal friction (Kozai-Lidov
cycles+tidal friction; KCTF) in such binaries has an important impact on the evolution of the binary components, leading to orbital shrinking and
even mergers \citep[a scaled up version of the KCTF in stellar triples;
e.g.][]{ 2006epbm.book.....E, 2007ApJ...669.1298F, 2009ApJ...697.1048P,
2009ApJ...699L..17P, 2012ApJ...747....4P, Prodan2013, Prodan2014, KatzDong2012, 2013MNRAS.431.2155N, 2013ApJ...773..187N, NaozFabrycky2014}. We follow the KCTF evolution of the binaries
until they merge or until at least one of the binary components evolves beyond
the MS, at which point we stop the full dynamical evolution of these binaries.Their later stellar evolution is then followed in isolation (the full coupling of secular
dynamical triple evolution with binary stellar evolution is beyond
the scope of this paper, and we only follow this limited, simplified
approach). Though highly simplified, this method allows us to track
for the first time the effects of the KCTF evolution during the MS
phase, and their implications for the long term stellar evolution
of NSC binaries.

  In Section \ref{sec:dynamics} we describe the
KL dynamics in the presence of additional forces and dissipation due to tides, and describe the relevant timescales. The choices of the initial conditions and the binary stellar evolution parameters are discussed in Section \ref{sec:initial}. In Section \ref{sec:numerics} we describe the results of numerical integration of the equations of motion using both orbit-averaged and N-body approach, as well as results of the binary stellar evolution. We discuss  our findings in Section \ref{sec:discussion} and summarize. 

\section{THREE BODY DYNAMICS IN THE PRESENCE OF ADDITIONAL FOCES} \label{sec:dynamics}

\subsection{The Kozai--Lidov mechanism}
We consider triple systems in hierarchical configurations, i.e. system in which the ratio between the outer binary semi-major axis (SMA) and the inner binary SMA is large. In such systems even small gravitational perturbations by the outer third body can significantly affect the inner binary orbital evolution on long enough secular timescales. The changes in the orbital elements of the inner binary can be particularly dramatic when the mutual inclination between the two orbits is high. Such configuration leads to the exchange of angular momentum between the inner and the outer orbits, resulting in periodic oscillations in the eccentricity of the inner orbit as well as the mutual inclination. The critical mutual inclination for having these oscillations, known as KL cycles  \citep{1962AJ.....67..591K}, is $i_{crit} \approx 39.2^{o}$ (for initially circular orbits of the inner binary). If the orbits are prograde ($i_{crit}\leq i\leq90^{o}$) these cycles are out of phase: when the eccentricity reaches its maximum, the mutual inclination reaches its minimum and vice versa. If the orbits are retrograde ($i>90^{o}$) these cycles are in phase: both the eccentricity and the mutual inclination reach maximum values simultaneously. The period of a KL cycle are much longer then the period of both the inner and the outer binary orbits in the triple. Such long term evolution can therefore be modelled using the secular approximation. In this approximation the equations of motion are averaged over the orbital periods of the inner and the outer binary. In the orbit averaged equations only the exchange of angular momentum between the inner and outer orbits is possible, but energy is not exchanged. Hence, such secular evolution can not, by itself,  cause changes in the SMA of either of the two orbits. 

When the separation between the stars in the inner binary  becomes sufficiently small, other physical processes beside pure Newtonian gravitational dynamics come into play.  In this work we consider several such effects that can induce additional periapse precession and thereby couple to the KL dynamics and typically tend to suppress it. These additional sources of precession include: apsidal precession due to tidal and rotational bulges, apsidal precession due to general relativity (GR) and the apsidal precession due to tidal dissipation, which is negligible in comparison to the former processes. 

The precession rate due to KL mechanism can be either positive or negative. In contrast, the precession rates due to GR and the tidal bulge effect are always positive and tend to promote periapse precession. As a consequence of the interplay between Kozai precession and GR and tidal bulge precession, the maximum eccentricity attainable by the system is lower; at the same time, the critical inclination at which KL evolution becomes significant increases (\citet{2001ApJ...562.1012E, 2002ApJ...576..894M, 2003ApJ...589..605W, 2007ApJ...669.1298F}, see their Figure 3). Precession rate caused by the rotational bulge may have either positive or negative value. The precession rates due to rotational and tidal bulges raised on both of the stars in the inner binary are parametrized by the tidal Love number $k_2$, a dimensionless constant that relates the mass of the multipole moment (created by tidal forces on the spherical body) to the gravitational tidal field in which that same body is immersed. $k_2$ encodes information on the internal structure of the stars\footnote{Note that the apsidal precession constant, which is a factor of two smaller than the tidal Love number, but which we do not utilize, is often denoted by $k_2$ as well.} and since we consider main sequence stars we adopt $k_2= 0.028$, a value characteristic for Sun-like stars \citep{2001ApJ...562.1012E, 2007ApJ...669.1298F}.

When the separation between the stars in the inner binary is of the order of a few stellar radii, tidal dissipation due to a close periapse approach in an eccentric orbit, or due to an asynchronous rotation may play an important role in the dynamical evolution \citep{1979A&A....77..145M, 2001ApJ...562.1012E}. As the inner binary orbit goes through the phases of high eccentricity the periapse distance may become sufficiently small to induce strong tidal dissipation. The tidal dissipation drains the energy from the orbit, while conserving the angular momentum. As a result the SMA shrinks, which in turn leads to an even stronger dissipation. Since the angular momentum is conserved in this process, it also results in the decrease of orbital eccentricity. Eventually the orbit circularizes at a separation of only a few stellar radii. Tidal dissipation is parametrized by the tidal dissipation factor $Q$, defined as the ratio between the energy stored in the tidal bulge and the energy dissipated per orbit.  We adopt $Q=10^6$, typically considered to be the characteristic value for sun-like stars \citep{2001ApJ...562.1012E, 2007ApJ...669.1298F}.

\subsection{Timescales}\label{timescale}
The dense stellar environment of a galactic nucleus is prone to dynamical
processes which do not typically take place in the field. Such processes can 
significantly alter the dynamics and eventually the stellar evolution of binaries near the MBH.
In this section we briefly review the relevant dynamical processes
associated with NSCs hosting MBHs and their timescales.  We compare the latter
with the secular KL timescales as well as with the timescales associated with
the binaries themselves. Additional details on the timescale calculations for the processes considered here can be found in \citet{2012ApJ...757...27A}.

Let us consider a binary with stellar components of mass $m_1$ and $m_2$, orbiting a MBH of mass $M_{\bullet}$. The eccentricities of the inner and the outer binary are denoted by $e_1$ and $e_{out}$, and the SMAs are denoted by $a_1$ and $a_{out}$, respectively. The argument of the periapsis of the inner binary, $\omega_{in}$ is defined relative to the line of the ascending nodes, while $i$ is the mutual inclination between the inner and the outer orbit.

 Binaries in a dense environment such as the Galactic center are susceptible to evaporation due to dynamical interactions with the surrounding stars. The ratio of the kinetic energy of the field stars to the internal orbital energy of the binary  determines whether the binary will evaporate; if this ratio is larger than unity, binaries are expected to evaporate.
The evaporation timescale for soft binaries is  \citep{2008gady.book.....B}:
\begin{eqnarray}\label{eqn:Tev}
T_{\rm EV}&=&\frac{M_{\rm b}\sigma}{G16\sqrt{\pi} M a_1\rho  \ln\Lambda}\\ \nonumber
&\approx& 3\times10^8\left(\frac{2}{\ln\Lambda}\right)\left(\frac{M_b}{2M_{\odot}}\right)\left(\frac{M_{\odot}}{M}\right)\left(\frac{\sigma}{100km/s}\right)\\ \nonumber
&& \times\left(\frac{1AU}{a_1}\right)\left(\frac{\rho_0}{\rho}\right) ~{\rm yrs},
\end{eqnarray}
where   $r$ is the distance from the MBH, $\rho$ the local density of stars, ${\rm ln} \Lambda$  the Coulomb logarithm, $M$ the mass of the field stars, 
 $M_{\rm b}=m_1+m_2$, $\sigma=\sqrt{GM_{\bullet}/r}/(1+\gamma)$ is
the local 1D velocity dispersion and $\gamma$ the slope of the stellar density profile.
Hereafter, we adopt $M_{\bullet}= 4\times 10^6 M_{\odot}$ \citep{2008ApJ...689.1044G, 2009ApJ...692.1075G}. We use the values of the normalization parameters $\rho_0=~{\rm 5.2\times10^5[\frac{M_{\odot}}{\pc^3}]}$ and $r_0=0.5~{\rm pc}$ typical for a GC-like nucleus  and $M=1M_\odot$, and $\gamma=2$ -- the slope expected for a dynamically relaxed single-mass population around a MBH.

 Binaries orbiting the central MBH with a mutual inclination $i \gtrsim 40^\circ$ undergo KL periodic variations of their eccentricity and inclination on a timescale:
 \begin{eqnarray}
T_{Kozai} &\approx &\frac{4}{3\sqrt{G}}  \left(\frac{a_{out}}{a_1}\right)^3 \frac{M_b^{1/2}}{M_{\bullet}}a_1^{3/2} (1-e_{out}^2)^{3/2}\\ \nonumber
&\approx & 2.5\times 10^6\left(\frac{a_{out}}{0.5{\rm pc}}\right)^3\left(\frac{1AU}{a_1}\right)^3\left(\frac{M_b}{2M_{\odot}}\right)^{1/2}\\ \nonumber 
&& \times\left(\frac{4\times10^6M_{\odot}}{M_{\bullet}}\right)\left(\frac{a_1}{1AU}\right)^{3/2}(1-e_{out}^2)^{3/2}~{\rm yrs}.
 \end{eqnarray}
 
Given a simple power law density model, $\rho\sim r^{-\gamma}$, setting 
$T_{Kozai} \approx T_{\rm EV}$ and $r=a_{out}$ gives 
the radius below which binaries can undergo 
 at least one KL cycle before they evaporate  due to gravitational encounters with surrounding stars:
\begin{eqnarray}
\tilde{r}_{\rm EV}&=&\left(\frac{6.8\times 10^{-4}}{\left([1+\gamma]\rho_0 r_0^\gamma \right)^2} \frac{a_1}{\ln \Lambda^2}
\left(\frac{M_{\bullet}}{M_b}\right)^3 \left(\frac{M_b}{M} \right)^2 \right.\\ \nonumber &&\times\left. \frac{M_b^2}{(1-e_{out}^2)^{3}}  \right)^{1/(7-2\gamma)}  \\ \nonumber
& \approx& \frac{0.885}{(1-e_{out}^2)} \left(\frac{2M_{\odot}}{4\times 10^6 M_{\odot}}
 \frac{M_{\bullet}}{M_b} \right) \left(\frac{1}{2} \frac{M_b}{M_{\odot}} \right)^{4/3}\\ \nonumber &&\times\left( \frac{4}{\ln \Lambda^2} \frac{a_1}{\rm AU} \right)^{1/3}~{\rm pc},
 \end{eqnarray}

 For $a_{out} \gtrsim \tilde{r}_{\rm EV}$  the binary evaporates before completing one KL cycle. For realistic values of the adopted parameters,  $\tilde{r}_{\rm EV}$ is comparable to the extent of the disk of young stars at the Galactic center and
can be of order the SgrA* influence radius.  However, at radii smaller than this critical radius KL cycles are detuned 
by fast relativistic precession of the inner binary orbit. It can be shown that general relativistic precession in the inner
binary suppresses the KL oscillations at radii larger than \citep{1997Natur.386..254H, 2002ApJ...578..775B}: 
\begin{eqnarray}
\tilde{r}_{\rm SC} & =& 0.02 \left(\frac{a_{1}}{{\rm AU}}\right)^{4/3}\left(\frac{M_{b}}{2M_{\odot}}\right)^{-2/3}\nonumber \\
&  & \times\left(\frac{M_{\bullet}}{4\times10^{6}{\rm M_{\odot}}}\right)^{1/3}\left(\frac{1-e_{1}^{2}}{1-e_{out}^{2}}\right)^{1/2} {\rm pc}~.\end{eqnarray}

For $a_{out}  \gtrsim \tilde{r}_{\rm SC}$ KL cycles are quenched by rapid Schwarzschild apsidal precession.
This preliminary analysis shows that the MBH induced KL cycles on GC binaries might be only important 
inside a tenth or possibly up to a few tenths of a parsec from the center (where the exact distance depends on the density profile model and orbital eccentricity). The maximum eccentricity such binaries can attain is not largely affected by
 relativistic precession while, at the same time, they can perform several KL oscillations before being dissociated 
 by encounters with field stars or before their components evolve to leave the main sequence.

\section{INITIAL CONDITIONS and methodology}\label{sec:initial}

On the basis of  the previous discussion, we concluded that only binaries on orbits passing relatively close to the MBH can be
significantly affected by KL oscillations. Therefore, in the following we only consider binaries within a galactocentric 
radius  $\lesssim 0.5 \pc$;  this radius also corresponds roughly to the outer extent of the young stellar disk in the central parsec of the GC.
The inner parsec of the Galaxy contains two distinct populations of young stars. One population, mainly  O  giants/supergiants and WR stars, is observed to have a disk-like structure extending from $0.5~$pc inward to within $0.05~{\rm pc}$ of the MBH~\citep{2009ApJ...697.1741B,2013ApJ...764..155L}. A second population of young stars consists of longer lived B-type main sequence stars that appear more isotropically distributed \citep{2010ApJ...708..834B,per+10}. The B-stars with projected radii of less than one arcsecond are usually referred to as the ``S-stars''. 
 We consider binaries originating from both the stellar disk and the stellar cusp
surrounding it. The total number of integrated binaries is 1367 for those originating from the disk, and 1670 for those originating from the cusp. In Section \ref{sec:inner_binary} we describe the choice of parameters for the binaries, while in Section \ref{sec:outer_binary} we discuss the choice of parameters for the orbits of the binaries around the MBH, i.e. the distribution of the outer orbits of the MBH+binary triples.

\subsection{Inner binary parameters}\label{sec:inner_binary}


\begin{table*}
\begin{centering}
\begin{tabular}{l|l|l}
\tableline
\multicolumn{3}{c}{TABLE 1. Inner binary parameters} \\
\tableline
\tableline
Symbol & Definition & Distribution \\ \hline
$m_1$ & primary mass & IMF with $\alpha_{DISK}=1.7$ and  $\alpha_{CUSP}=2.35$\\
$m_2$ & secondary mass & $40\%$ twins, $60\%$ $(m_2/m_1)$ uniform in (0,1) \\
$a_1$ & Inner binary semimajor axis & lognormal with  $<logP(d)>=4.8$  and $\sigma(d)=2.8$\\
$e_{1, 0}$ & Inner binary initial eccentricity & thermal, $e_{in,0}<0.9$ \\
$i_{init}$ & Initial mutual inclination & uniform in $\cos(i)$ \\
$\omega_{in, 0}$ & Initial argument of periastron & uniform\\
$\Omega_{in}$ & Longitude of ascending node & uniform\\
$R_{1,2}$ & stellar radius & $R_{1,2}=(m_{1,2}/M_{\bigodot})^{0.75}R_{\bigodot}$\\
$r_{p,in}$& inner binary periapse& $r_{p,in}\geq5(R_1+R_2)$\\
$k_2$ & Tidal Love number & $0.028$\\
$Q$ & Tidal dissipation factor & $10^6$\\
\tableline
\tableline
\end{tabular}
\par
\end{centering}
\end{table*}



Binaries originating from the stellar disk are assumed to be relatively massive;  the masses of primary stars, $m_1$, are drawn from a top heavy initial mass function (IMF) with a power-law slope of  $\alpha_{DISK}=1.7$ \citep{2013ApJ...764..155L}. The primaries of binaries originating from the stellar cusp are assumed to follow a Salpeter IMF with slope of $\alpha_{CUSP}=2.35$ \citep{1955ApJ...121..161S}. We set the mass of the secondary stars $m_2$ to be equal to $m_1$ in $40\%$ of the cases, while the others are chosen by selecting the mass ratio $q_{in}=m_2/m_1$ uniformly between $0$ and $1$. The stars are initially taken as MS stars with appropriate MS radii, $R_{1,2}=R_{\odot}(m_{1,2}/M_{\odot})^{0.75}$. The initial orbital period of the (inner stellar) binaries is chosen from a log-normal distribution following \citet{1991A&A...248..485D}, with $\langle log P(days) \rangle=4.8$ and $\sigma_{log P(days)}=2.8$. A thermal distribution for the inner binary eccentricity is assumed. The initial mutual inclination between the binary orbit and the orbit around the SMBH is uniform in $\cos i$. The argument of periapse and the longitude of the ascending node are uniformly distributed. For tidal Love number, tidal dissipation factor, and moment of inertia of the binary components we use values of $k_2=0.028$, $Q=10^6$ and $I_{1,2}=0.08m_{1,2}R_{1,2}^2$, respectively, which are typical for Sun-like stars \citep{2001ApJ...562.1012E}. The initial spin periods of the binary components is set to $10$ days and the spin angular momentum is aligned with the orbital angular momentum of the binary. We only integrate systems for which the initial inner binary periapse is larger than $5\times (R_1+R_2)$, to assure that the outcome of the integration is indeed due to KL mechanism and not pure tidal dissipation.  Table 1 summarizes our choice of parameters for the inner binary orbits.

\subsection{Outer binary parameters}\label{sec:outer_binary}

To model the mass density of stars in an NSC similar to the GC we use a power-law density profile:
\be\label{eq:rho}
\rho(r) = \rho_0\left(\frac{r}{r_0}\right)^{-\gamma}\left[1 + \left(\frac{r}{r_0}\right)^2\right]^{(\gamma-1.8)/2},
\ee
where here $\gamma$ is the slope of the inner density profile. We adopt $r_0=0.5\pc$ and $\rho_0=5.2\times10^5[\frac{M_{\odot}}{\pc^3}]$, which give a good fit to the observed spatial density at the GC outside the core (normalized at $1\pc$; \citep{2009A&A...502...91S}). All timescales considered are computed for two models for the inner density profile: (1) a shallow power-law slope, $\gamma=0.5$,  representing the observed distribution of stars at the GC \citep{2009A&A...499..483B, 2009ApJ...703.1323D, 2010ApJ...708..834B};  (2) a steep power-law slope, $\gamma \approx 2$, corresponding to a nearly relaxed configuration of stars in a potential dominated by a MBH \citep{2005PhR...419...65A}. 
The outer binary parameters depend on the origin of the binaries;
the stellar disk, that extends from $\sim0.04\pc$ to $\sim4\pc$,
or the stellar cusp.
 For each set of binaries we compute the binary evaporation time in this collisional environment considering  both the shallow density profile model  and the the steep cusp  density profile mentioned above. 
 For binaries originating from the stellar disk we draw the SMAs of the outer binaries (i.e, for the orbit around the MBH) from the distribution  $dN(a)/da \sim a_{out}^{-1}$ following \citet{2009ApJ...690.1463L} while for the eccentricity of the outer orbit we adopt the double-peaked distribution from \citet{2009ApJ...697.1741B} which has two maxima at $e_{out}\sim 0.35$ and $e_{out}\sim 0.95$.

 For binaries originating from the stellar cusp we sampled the orbital elements of the binary around the MBH from the following distribution:
 \be
 N(a,e^2) = N_0a^{2-\gamma}dade^2,
 \ee
 i.e. assuming a steady-state phase-space distribution for an isotropic cusp in the neighbourhood of a dominating point mass potential. Therefore, the SMAs are drawn from a $dN/da\sim a^{2-\gamma}$ distribution, while the eccentricity distribution of the binary orbit around the MBH is taken to be thermal. The background cusp determining the binary evaporation time was modelled using the density profile of equation~\ref{eq:rho}.
 
 For the shallow cusp density profile, ($\gamma=0.5$), we use equation \ref{eq:rho} where we set  $r_0=0.5\pc$ and $\rho_0=5.2\times10^5[\frac{M_{\odot}}{\pc^3}]$. We compute the evaporation time scale of the binary systems assuming stellar mass perturbers for the $\gamma=0.5$ model. In the steep cusp density profile model we compute the evaporation time taking into account both stellar and BH perturbers.  The combined density profile corresponds to a mass segregated cusp near a MBH :
 
 \be
 \rho(r)=\rho_{\star}(r) + \rho_{BH,0}\left(\frac{r}{0.5\pc}\right)^{-2},
 \ee 

where $\rho_{BH,0}=10^4[\frac{M_{\odot}}{\pc^3}]$ and {\bf $\rho_{\star}$} corresponds to the stellar density profile given by equation~\ref{eq:rho} with $\gamma=1.5$, the density profile slope of the main sequence star population in
the quasi-steady state multimass models of \citet{2006ApJ...645L.133H}.


\begin{table}[h]
\begin{center}
\begin{tabular}{l|l|l}
\tableline
\multicolumn{3}{c}{TABLE 2. Parameters determined by the model} \\
\tableline
\tableline
Symbol & Disk, $\gamma=1.5$ & Disk, $\gamma=0.5$\\ \hline
$a_{out}$ &$\frac{dN(a)}{da}\sim a_{out}^{-1}$ (Lu et. al 2009) &$\frac{dN(a)}{da}\sim a_{out}^{-1}$ (Lu et. al 2009) \\
$e_{out}$ & from Bartko et. al 2009.&from Bartko et. al 2009  \\
$\rho_{\star}$ [$\frac{M_{\bigodot}}{pc^3}$]  & $5.2\times10^5(\frac{a_{out}}{0.5pc})^{-1.5}$&$5.2\times10^5(\frac{a_{out}}{0.5pc})^{-0.5}$\\
\tableline
\tableline
Symbol & Cusp, $\gamma=1.5$ & Cusp, $\gamma=0.5$\\ \hline
$a_{out}$ &$\frac{dN(a)}{da}\sim a_{out}^{0.5}$ &$\frac{dN(a)}{da}\sim a_{out}^{1.5}$ \\
$e_{out}$ & thermal & thermal  \\
$\rho_{\star}$ [$\frac{M_{\bigodot}}{pc^3}$]  & $5.2\times10^5(\frac{a_{out}}{0.5pc})^{-1.5}$&
$5.2\times10^5(\frac{a_{out}}{0.5pc})^{-0.5}$\\
$\rho_{BH}$ [$\frac{M_{\bigodot}}{pc^3}$]  & $10^4(\frac{r}{0.5pc})^{-2}$&\\
\tableline
\tableline
$r_{p,out}$ & $r_{p,out}\geq 4\times r_{bt}$& $r_{p,out}\geq 4\times r_{bt}$\\
\tableline
\tableline
\end{tabular}
\end{center}
\end{table}



\subsection{Dynamical model: Secular KL evolution with
tidal friction (KCTF) in octupole approximation}\label{sub:KCTF}

We treat the gravitational effects of the third body in the octupole approximation, where we derive the equations of motion from the double-averaged Hamiltonian. In other words we average  over the orbital
periods of both the inner binary and the binary orbit around the MBH, and retain terms up to $(a_{1}/a_{out})$ to $3^{rd}$ order. Beside the perturbations due to the presence of the third body via KL mechanism, we include
the following dynamical effects:
\begin{itemize}
\item periastron advance due to general relativity in the inner binary;
\item periastron advance arising from quadrupole distortions of the inner binary stars due to both tides and rotation;
\item orbital decay due to tidal dissipation in the inner binary stars.
\end{itemize}
\noindent The equations used in our model are those of \citet{2000ApJ...535..385F} and \citet{2002ApJ...578..775B} for the octupole terms, combined with equations from \citet{2012ApJ...747....4P} for tidal effects on both stars in the binary. 
Since we integrate only systems for which $r_{p,out} > 4\times r_{bt}$, where $r_{bt}\sim \left(\frac{M_{\bullet}}{m_1+m_2}\right)^{1/3}a_{1}$, the secular approximation is justified. In this parameter regime the results of the octupole integration are in good agreement with the results of direct N-body integration \citep{2010ApJ...713...90A, 2012ApJ...757...27A}. During the integration of the binaries evolution we regard an event as a \textquotedbl{}merger\textquotedbl{} when one of the binary members starts overflowing it's Roche lobe, which for MS stars occurs at  $r_{p,in}\approx (R_1+R_2)$. Our code does not treat mass transfer or mass loss. The stopping condition for the integration are:
 \begin{itemize}
 \item binary \textquotedbl{}mergers\textquotedbl{} occurred,  $r_{p,in}\approx (R_1+R_2)$ or one of the binary members starts overflowing its Roche lobe;
 \item binary has evaporated due to dynamical interactions with field stars,   $t\approx T_{EV}$ ;
 \item one or both components reached the end of their main sequence lifetime $t\approx T_{MS}$ ;
 \item the maximum integration time for the  binaries originating from the stellar disk is $T_{max}=10^7\yr$ which corresponds approximately to the lifetime of the stellar disk; while the  maximum integration time for the  binaries originating from the cusp is set to $T_{max} =10^{10}\yr$.
 \end{itemize} 

\subsection{Binary stellar evolution}\label{sub:BSE}

We account for binary mergers due to KCTF during the MS lifetime of binary components in the dynamical evolution phase discussed above.
However, KCTF evolution could significantly affect the orbital evolution
of a binary without leading to a merger during the MS phase, and therefore
strongly impact its long term evolution beyond the MS. In
order to explore such possible effects, we follow the binary stellar
evolution of the non-merged binaries from the KCTF stage. We use
a simplified approach, in which we take the final orbital configuration
of all non-merged binaries from the KCTF stage, and then follow their
evolution using the binary stellar evolution population
synthesis code BSE developed by \citet{2000MNRAS.315..543H}. In order to
explore the importance of the role that KCTF plays during the MS phase on the binary stellar evolution, we also consider the evolution of the same binaries
using their initial configuration, before any KCTF evolution occurred.
In both cases we initialize the systems with zero age MS components.
Finally, we compare the results of the stellar evolution of both groups. 

 Such comparison provides a fraction of systems that are significantly affected
by KCTF, and are defined as concurring with at least one of the following:
\begin{enumerate}
\item The stellar type (as defined in the BSE code; e.g. a WD vs. a red-giant) of at least one of the stellar components in the KCTF evolved binary differs from its corresponding non-KCTF evolved binary. 
\item At least one of the binary components mass differs from its corresponding
non-KCTF evolved binary by more than 5\% (the fraction is with respect to the more massive relevant star, either from the KCTF or non-KCTF evolved case). 
\item The binary orbital period changed by at least 10 \%.
\end{enumerate}

The change in the eccentricity of the inner binary is a natural consequence of KL
evolution, but just on its own it might not notably impact the binary evolution.
Therefore, we do not consider this change in eccentricity by itself
 as a driver defining a significant difference in the evolution of the two groups. 
 
 There are two main caveats with this simplistic approach. First, during
the BSE phase we do not account for any KCTF evolution, while such
effects could be significant and further affect the binary evolution;
it is therefore likely that our calculated fractions of systems affected
by the combined KCTF and further BSE are only underestimates for the
more significant actual impact. Second, the time during which the
KCTF evolution was followed is not accounted in the later BSE, i.e.
the KCTF binaries are assumed to be zero age MS binaries, with only
their orbital parameters changed due to KCTF. For these reasons the
detailed types of binary systems produced in the KCTF+BSE (e.g. X-ray
binaries, CVs etc.) are not discussed; rather, only the overall fractions
of systems in which early KCTF evolution gave rise to significant
difference in the later BSE are calculated, providing a first order
estimate of the effects discussed in this paper. More detailed results
of population synthesis of binaries near MBHs is beyond the scope
of this paper, and would require a triple stellar evolution code,
in which binary stellar evolution and the triple system dynamics are
inherently coupled \citep[See][for
initial steps in this direction]{2012ApJ...760...99P, 2013MNRAS.430.2262H}.

\section{RESULTS}\label{sec:numerics}

\subsection{KCTF evolution}\label{sec:ktcf}

In this section we present the results of our numerical model for the secular evolution of the binaries around the MBH. As described in the previous section, we consider binaries originating from both the stellar disk and the stellar cusp. In both cases we 
model the background cusp considering either the shallow or the  steep density profile models in order to determine the relevant evaporation timescales. Details on how we generate the initial conditions 
are found in Section~\ref{sec:initial}. The outcomes of such evolution can be divided in following categories:
\begin{itemize}
\item \textquotedbl{}mergers\textquotedbl{}-- one of the binary components starts overflowing its Roche lobe or they physically collide ($r_{p,in}\approx (R_1+R_2)$). In the case of main sequence stars these two conditions are almost equivalents since main sequence stars need to be almost touching each other in order to start Roche lobe overflow. 
\item Evaporated binaries -- the two components of the binary are no longer bound together due to the interaction with the field stars. Whether the binary will evaporate or not strongly depends on the assumed density profile of the 
background cusp   and the binary SMA. For example, using
a steeper density profile increases the central density of background stars, reduces the binaries evaporation times
and in turn increases the percentage of evaporated systems.
\item Tidally affected binaries -- as long as the perturbation from the MBH effectively excites the eccentricity, the binary SMA may suffer dramatic shrinkage due to tidal dissipation. As the binary becomes tighter the probability 
for its evaporation decreases (see equation \ref{eqn:Tev}). 

\end{itemize}

\begin{figure}
\epsscale{1.0} 
\vspace{5mm}
\plotone{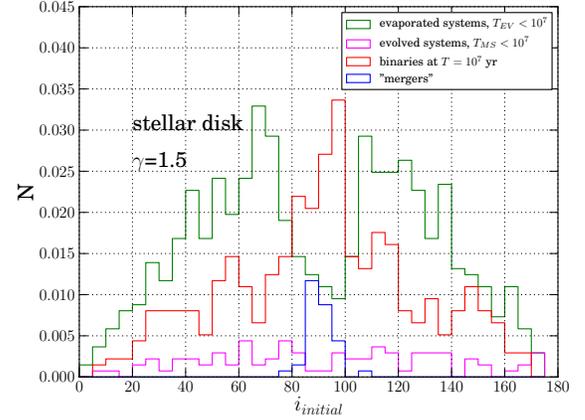}
\caption{Histogram of the initial inclination for the binaries in the stellar disk for  $\gamma=1.5$. Approximately $3\%$ of the systems are likely to merge while on the main sequence. All \textquotedbl{}mergers\textquotedbl{} on the main sequence occur in systems with high mutual inclinations and already during the first Kozai--Lidov cycle. The percentage of \textquotedbl{}mergers\textquotedbl{} is not affected by the choice of a density profile, but the number of systems that evaporate is higher when a steep density profile is considered, as expected.  \label{Fig:inc_disk}}
\end{figure}

\begin{figure}
\epsscale{1.0} 
\plotone{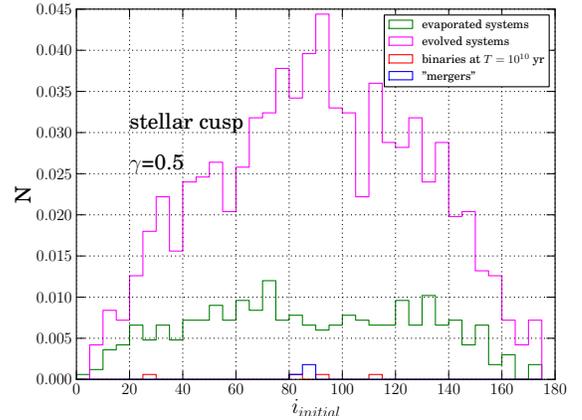}
\vspace{5mm}
\plotone{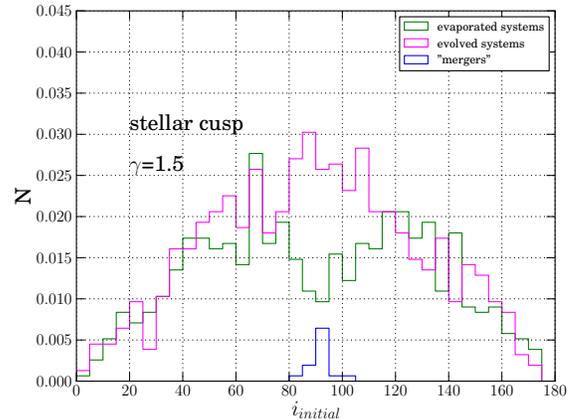}
\caption{ Histogram of the initial inclination of cusp binaries: left panel shows results for  $\gamma=0.5$ and the right panel shows results for $\gamma=1.5$. Approximately $2-3\%$ of the systems are likely to merge while on the main sequence. Similar to the case of stellar-disk binaries, all \textquotedbl{}mergers\textquotedbl{} occur for high mutual inclinations and already on the first Kozai--Lidov cycle. Again, the choice of the density profile does not affect the percentage of \textquotedbl{}mergers\textquotedbl{} and the number of evaporated systems is higher in the steep density profile model case, as expected.  \label{Fig:inc_cusp}}
\end{figure}

Figs. \ref{Fig:inc_disk} and \ref{Fig:inc_cusp} show the histograms for the binary evolution outcomes for both the stellar-disk binaries and the cusp binaries, and for both the shallow and the steep density profile models.  We find that $\sim 3\%$ of the stellar-disk binaries experience  ``mergers" while on the MS. The same is true for $\sim1\%$ of the steep cusp binaries and 
$\sim0.25\%$  of the shallow cusp model binaries  (see Table 3 as well). The majority of ``mergers" occur for initially highly inclined orbits and at the first maximum in the KL cycle. The merging systems are not very sensitive to evaporation precesses 
as the KL timescale for most of these systems is orders of magnitude shorter than their evaporation time (see Section~\ref{timescale}).
Due to the larger number of objects closer to the MBH, where $a_{out} \lesssim \tilde{r}_{\rm SC}$, the binaries in the steep cusp model
can in average achieve higher eccentricities and are more likely to merge than binaries in the shallow cusp model. As a consequence of 
this the number of mergers in the steep cusp model is about four  times larger than in the core model.

 The difference in the percentage of "mergers" between stellar-disk and cusp binaries is due to the selected IMF. As mentioned before, the mass of the binary components of the stellar-disk binaries is drawn from a top heavy IMF. Therefore, the ratio of their physical size to the separation is larger than that of the less massive cusp binaries (following a Salpeter IMF), resulting in a higher \textquotedbl{}merger\textquotedbl{} rate. The observed peak at high inclinations in the distribution of surviving stellar-disk binaries is due to those binaries that shrank significantly due to KCTF.
Such evolution also increases their evaporation time beyond the lifetime of the stellar disk ($10^7\yr$). The same effect produces a similar peak at high inclinations for cusp binaries followed until they evolved off the main sequence.
The shape of the density profile determines the percentage of evaporated binaries; when a steep density profile model is assumed, the number/percentage of evaporated systems is higher, as expected. On that account, at the end of the calculated dynamical evolution, after $10^7\yr$, the distributions of the surviving systems in the shallow and steep density profile models is distinctively different. 

%
%
%

\begin{figure}
\epsscale{1.0} 
\vspace{5mm}
\plotone{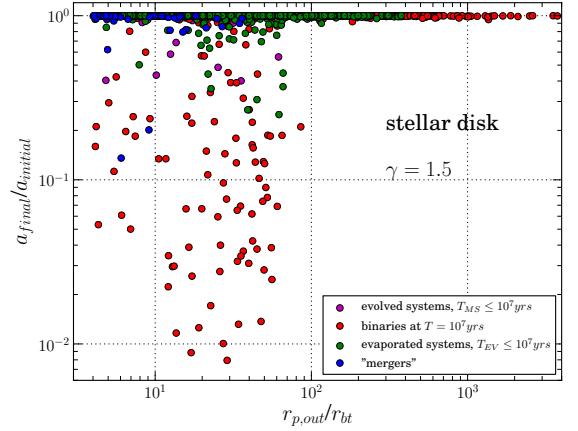}
\caption{$a_{final}/a_{initial}$ vs. $r_{p,out}/r_{bt}$ for the stellar disk: $\gamma=0.5$ (upper panel) and $\gamma=1.5$ (lower panel). Both choices of $\gamma$ demonstrate that  for the majority of \textquotedbl{}mergers\textquotedbl{} tidal dissipation is not relevant ($a_{final}/a_{initial} \sim1$).  The region where tidal dissipation is important is $r_{p,out} < 100\times r_{bt}$. Once the stars leave the main sequence, their radius expands due to stellar evolution, and tidal effects may become even stronger. Such a combination of tidal effects and Kozai--Lidov cycles could lead to post main sequence coalescence/ strong binary interaction.
  \label{Fig:semi_disk}}
\end{figure}

\begin{figure}
\epsscale{1.0} 
\vspace{5mm}
\plotone{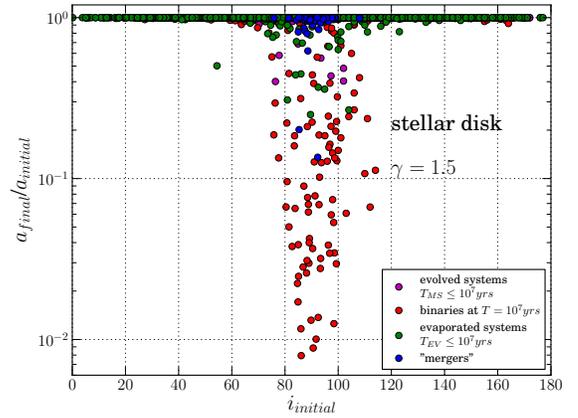}
\caption{$a_{final}/a_{initial}$ vs. $i_{initial}$ for the stellar disk for $\gamma=1.5$. For the majority of \textquotedbl{}mergers\textquotedbl{} tidal dissipation is not relevant.  Tidal dissipation is important for inclinations in the range $70^{o}\lesssim i_{initial}\lesssim110^{o}$. Similar results are obtained for the cusp model.
  \label{Fig:a0af_disk_inc}}
\end{figure}

Fig. \ref{Fig:semi_disk} shows the dependence of the ratio of the final SMA  the initial SMA of the inner binary on the closest approach to the MBH, scaled to the tidal disruption radius; ($a_{final}/a_{initial}$ vs. $r_{p,out}/r_{bt}$). The stellar-disk  model clearly demonstrates that for the majority of \textquotedbl{}mergers\textquotedbl{} tidal dissipation is not important. In other words,  the KL timescale is much shorter than the tidal dissipation timescale due to the strong Kozai torque ($a_{final}/a_{initial} \sim1$), and the same is true for the stellar cusp model. Since the period of KL cycles strongly depends on the separation of the perturber  ($T_{Kozai}\sim a_{out}^3$), the region where tidal dissipation becomes important is  $r_{p,out} < 100\times r_{bt}$.  For many binaries in this region, Kozai torque is strong enough to induce significant eccentricity oscillations but the eccentricity does not reach sufficiently high values to result in a  \textquotedbl{}merger\textquotedbl{}. Instead, during the high
eccentricity phase of a KL cycle, the periapse separation of the binary becomes small enough for tides to 
become important to the evolution. As seen in Figure \ref{Fig:a0af_disk_inc} the systems that experience dramatic tidal evolution are those in the KL high inclination regime. Due to the energy drained by tidal dissipation at periapse passage, their SMA significantly shrinks. After these stars evolve off the MS, their radii expand due to stellar evolution which, in principle, could result in an even stronger tidal effects. Eventually, such combination of KL cycles with tidal friction may lead to the binary coalescence or strong binary interaction during the post-MS phase. 

The shrinkage of the SMA of the inner binary due to KCTF becomes important when it comes to the survival of  binaries in the environment of the GC. Our calculations demonstrate that the survival
rate of the binaries is $\sim10\%$ higher than in the case where
KL evolution is neglected (due to the binary hardening as a result
of KCTF-induced SMA shrinkage).

\begin{figure}
\epsscale{1.0} 
\vspace{5mm}
\plotone{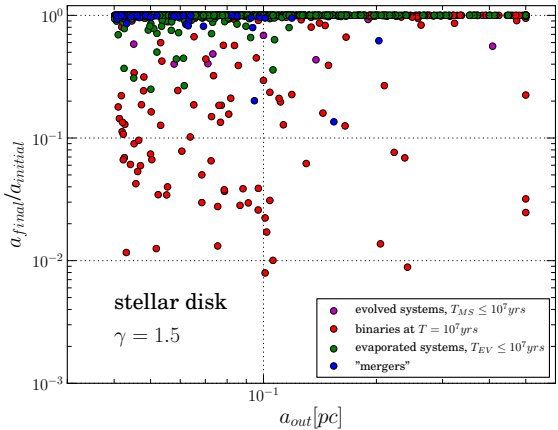}
\caption{$a_{final}/a_{initial}$ vs. $a_{out}$ for the stellar disk and a steep density profile model ($\gamma=1.5$). The majority of \textquotedbl{}mergers\textquotedbl{} occur within $0.1\pc$ off the MBH and are not driven by tidal dissipation ($a_{final}/a_{initial} \sim1$).  Systems in this region that did not coalesce were strongly affected by KCTF evolution. 
  \label{Fig:a0af_disk}}
\end{figure}

\begin{figure}
\epsscale{1.0}
\vspace{5mm} 
\plotone{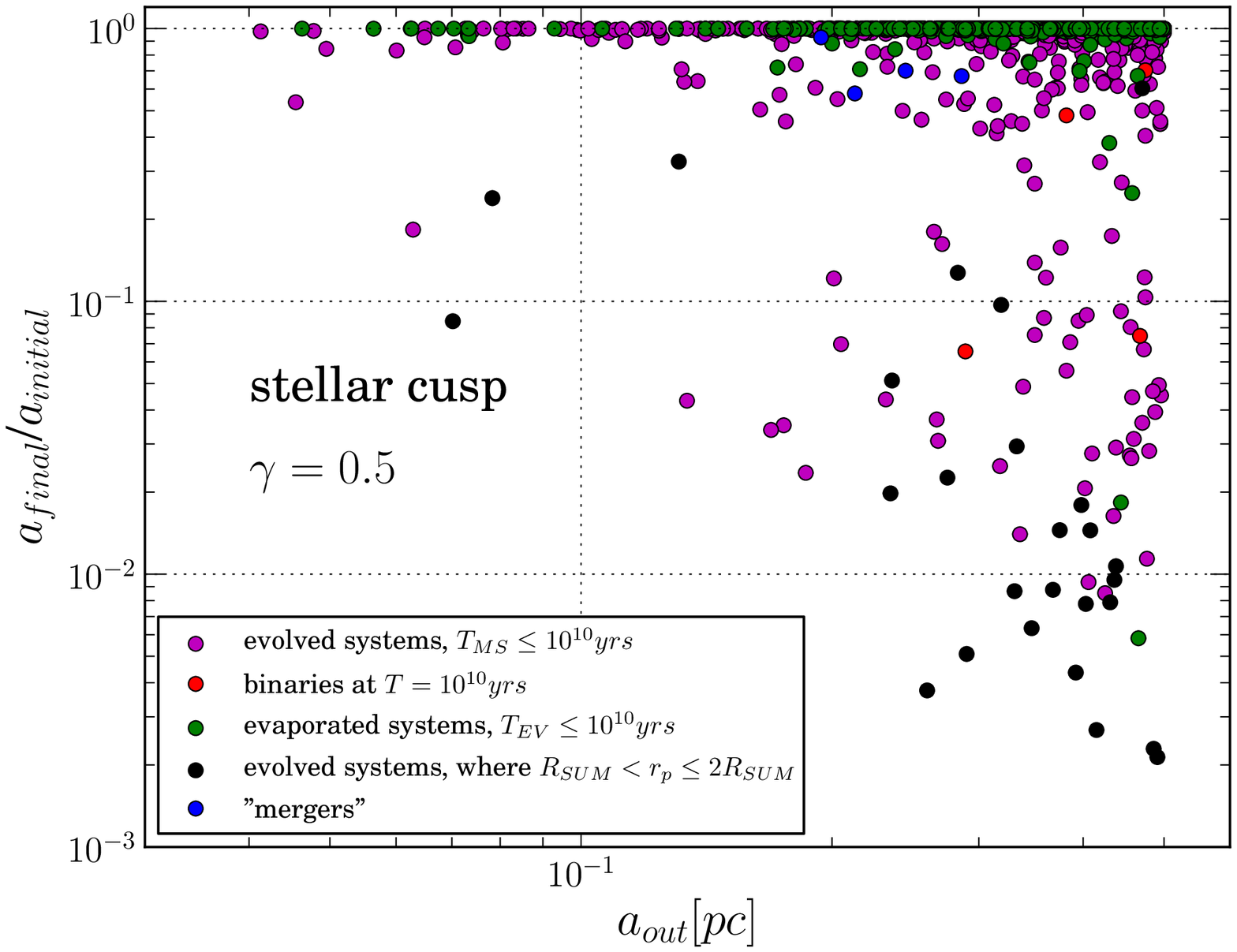}
\vspace{7mm}
\plotone{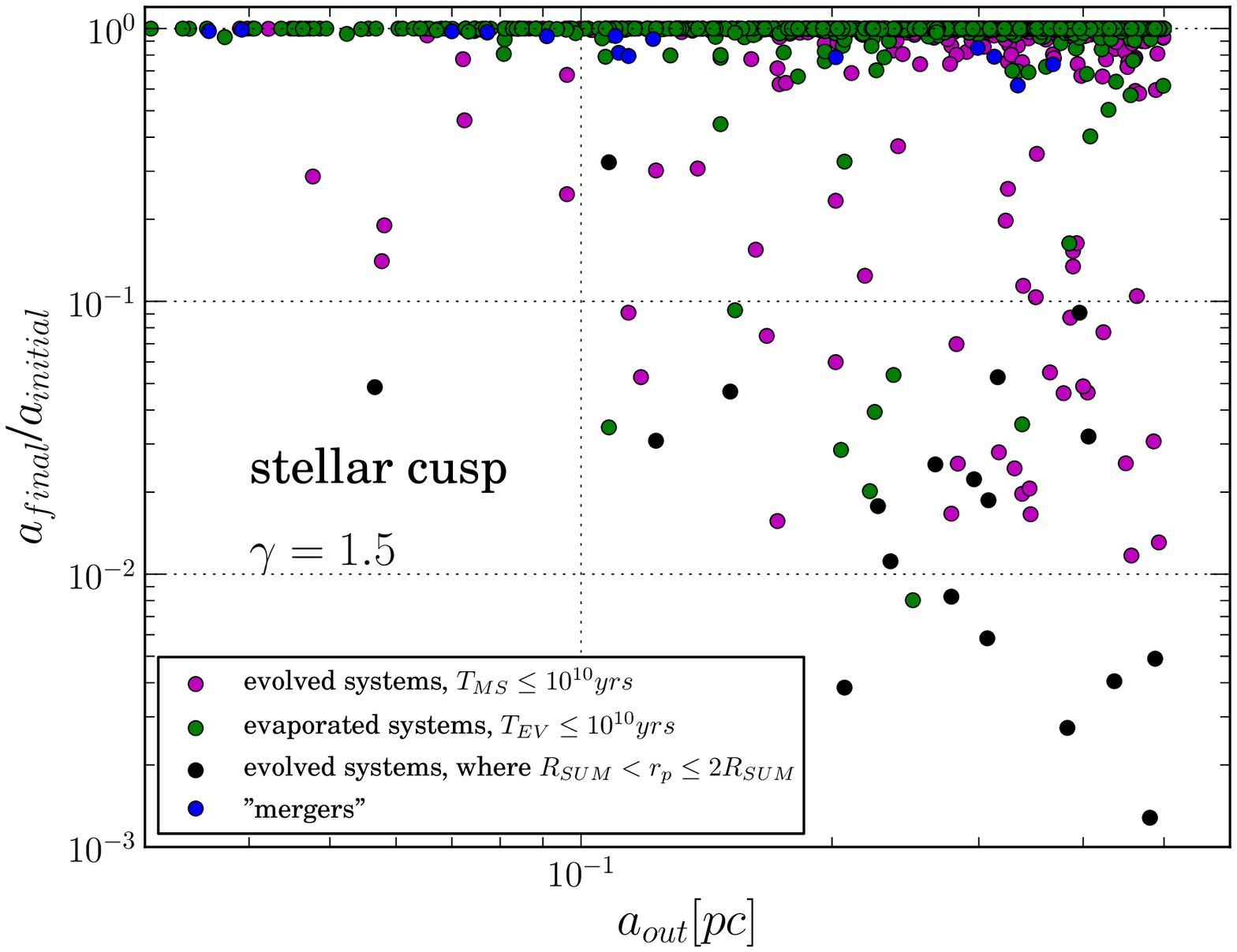}
\caption{$a_{final}/a_{initial}$ vs. $a_{out}$ for the stellar cusp and a the two density profile models ($\gamma=0.5$) (upper panel) and $\gamma=1.5$ (lower panel). Again, both models demonstrate that the majority of \textquotedbl{}mergers\textquotedbl{} are not driven by tidal dissipation ($a_{final}/a_{initial} \sim1$), and KCTF becomes important for systems in the central 0.1 pc region.
  \label{Fig:a0af_cusp}}
\end{figure}

Figs. \ref{Fig:a0af_disk} and \ref{Fig:a0af_cusp} depict the dependence of the ratio of the final SMA to the initial SMA of the inner binary, on the outer SMA (separation of the inner binary from the MBH) for the stellar-disk and cusp binaries, and for both the shallow and the steep density profile models. As already emphasized, both figures demonstrate that tidal dissipation is not a main driver for the occurrence of  \textquotedbl{}mergers\textquotedbl{}. Figure \ref{Fig:a0af_disk} shows that tidal dissipation significantly impacts systems within a distance of $\sim0.1\pc$ from the MBH. Such systems shrink and harden due to KCTF and can survive stellar scatterings longer; ($T_{EV}$ becomes longer as $a_{in}$ shrinks). This process can, in principle, affect the observed period distribution of stars in the stellar-disk.  Fig. \ref{Fig:a0af_cusp} shows that all of the tidally affected cusp binary systems reside beyond $\sim0.1\pc$ from the MBH. A number of these system has periapse distance of only few stellar radii i.e. systems for which $R_{SUM} < r_p \leq 2\times R_{SUM}$. The difference in the location of regions containing tidally affected systems in the disk and in the cusp lies in the choice of initial distribution of $a_{out}$ (see Table 2).

\begin{figure}
\epsscale{1.0} 
\vspace{5mm}
\plotone{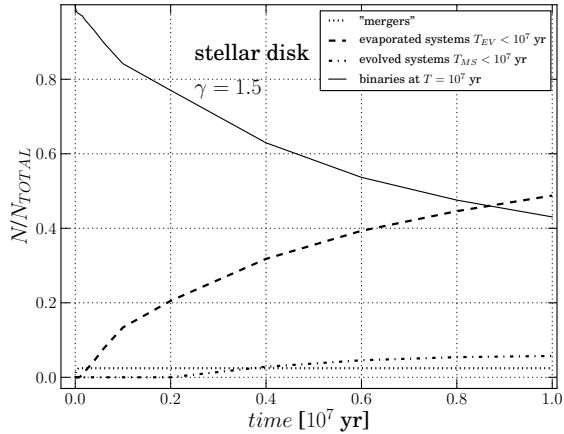}
\caption{Cumulative distribution of '\textquotedbl{}mergers\textquotedbl{}, evaporated systems, evolved systems and surviving binaries for the binaries in the stellar disk with $\gamma=1.5$. \label{Fig:cdf_disk}}
\end{figure}

\begin{figure}
\epsscale{1.0} 
\vspace{5mm}
\plotone{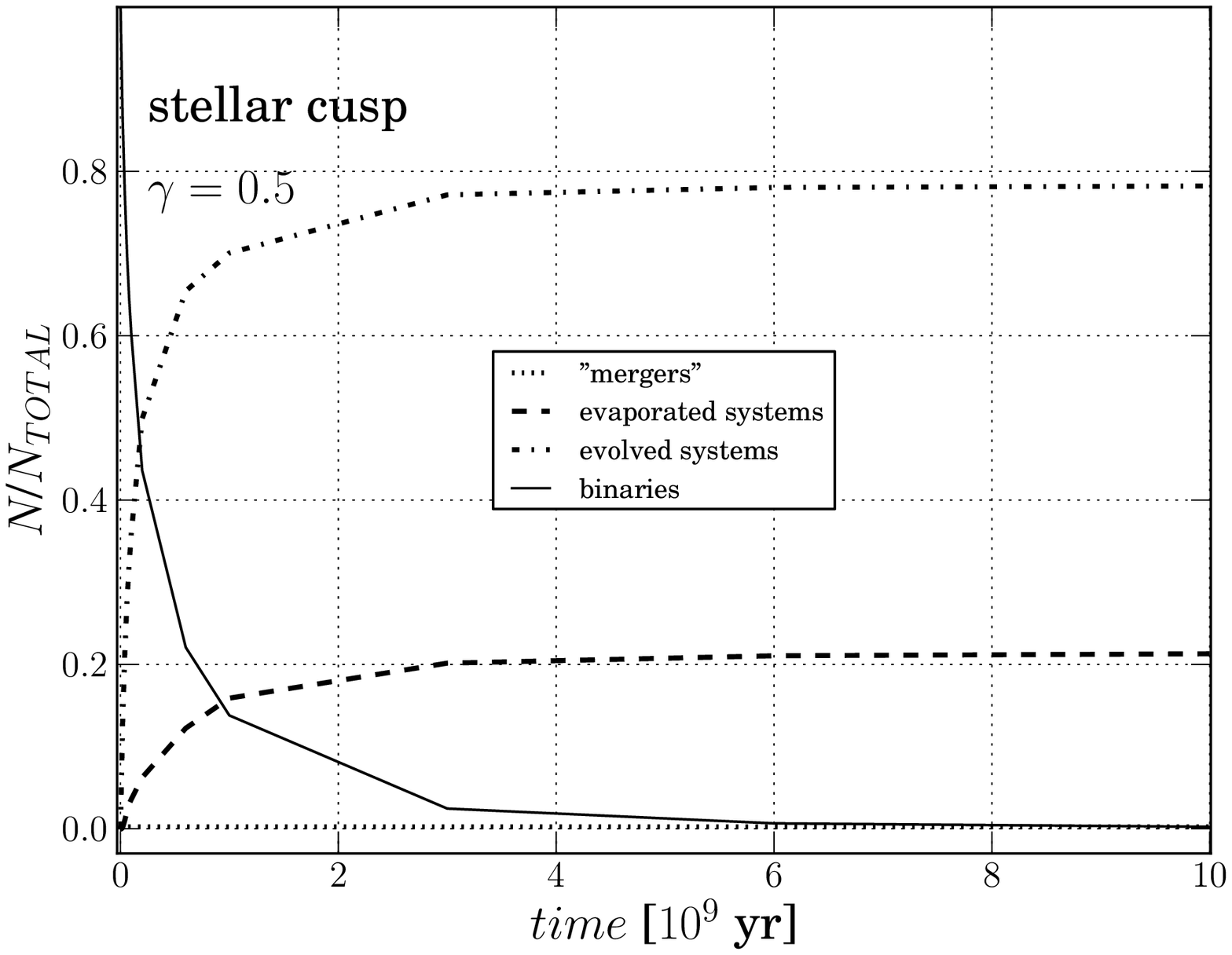}
\vspace{7mm}
\plotone{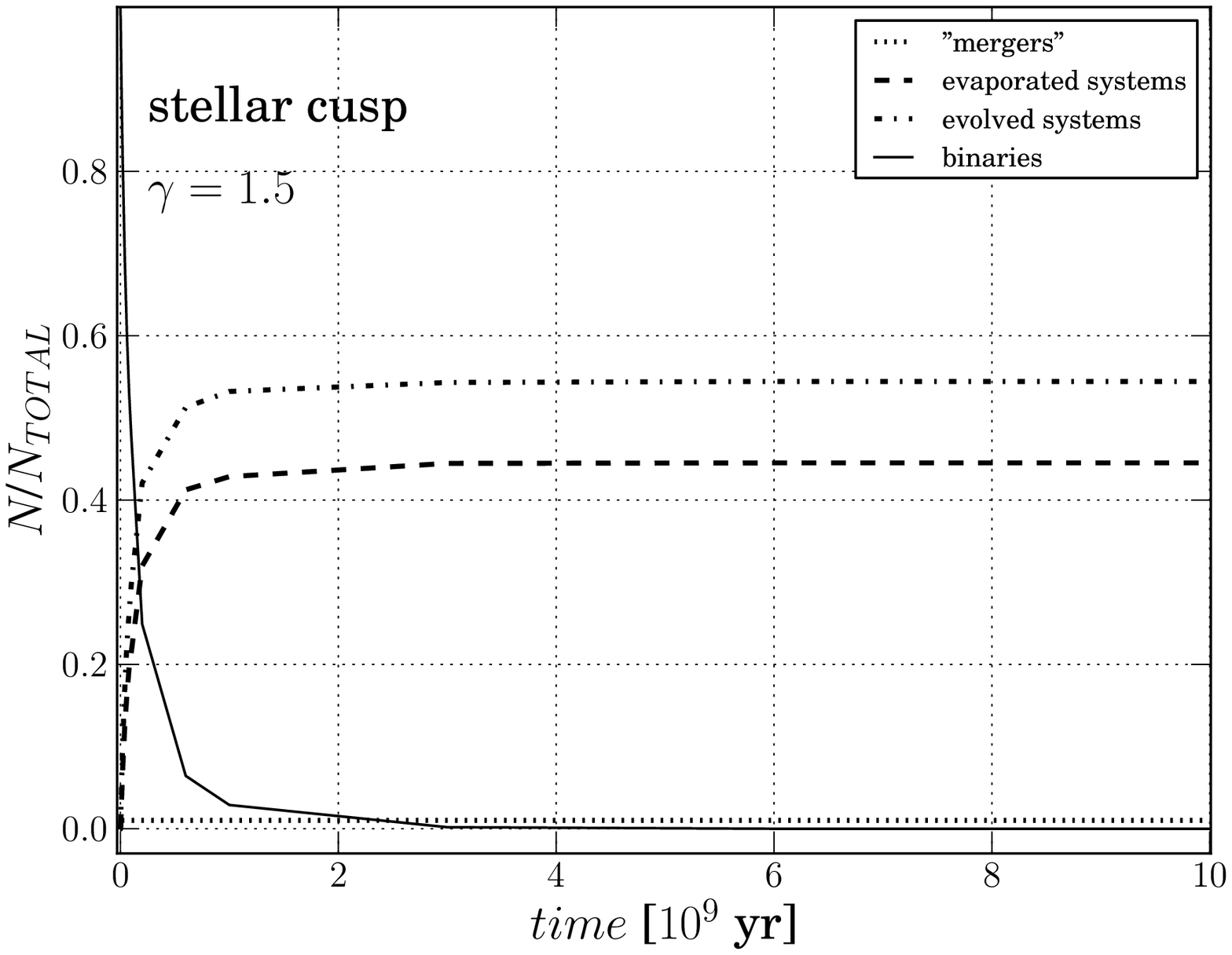}
\caption{ Cumulative distribution of  \textquotedbl{}mergers\textquotedbl{}, evaporated systems, evolved systems and surviving binaries for the binaries in the stellar cusp and the density profile models $\gamma=0.5$ (upper panel) and $\gamma = 1.5$ (lower panel). The fraction of  \textquotedbl{}mergers\textquotedbl{} is not affected by the choice of the slope of the density profile while the number of evaporated systems clearly is. Steeper density profile leads to higher fraction of evaporated systems.\label{Fig:cdf_cusp}}
\end{figure}

Figs. \ref{Fig:cdf_disk} and \ref{Fig:cdf_cusp} show the cumulative distribution of   \textquotedbl{}mergers\textquotedbl{}, evaporated systems, evolved systems and surviving binaries for both stellar-disk and cusp binaries. As seen in Fig.  \ref{Fig:cdf_cusp}, the
majority of mergers occur at early times during the first Kozai cycle.
 We found that the fraction of mergers is sensitively higher if the binaries originate in 
a stellar disk  characterized by a top heavy initial mass function. 
In this latter case  the 
ratio of the stars physical size to their separation in a binary is larger than that of the less massive cusp binaries,
which increases the chance for a close stellar interaction. 
The density profile is an important factor in determining the percentage of evaporated systems which, as expected, is significantly higher in the case of a steep density profile ($\gamma=1.5$) model.

All our integrations and their outcomes are summarized in Table 3.


\begin{table}
\begin{center}
\begin{tabular}{l|l|l|l|llllllllllllllllllllllllllllll}
\tableline
\multicolumn{5}{c}{TABLE 3. Results of simulations} \\
\tableline
\tableline
fraction of&Disk & Disk & Cusp &Cusp\\
systems &$\gamma=1.5$ & $\gamma=0.5$ & $\gamma=1.5$ &$\gamma=0.5$ \\ \hline
 ''mergers" & $0.03$ & $0.03$& $0.01$ & $0.0025$\\
evaporated  & $0.554$& $0.043$& $0.45$ & $0.213$\\
evolved  & $0.063$ & $0.073$ & $0.54$ & $0.782$\\
 binaries & $0.353$ & $0.85$& $0$& $0.0025$\\
\tableline
\tableline
integration &&&&\\
$T_{max}$ [yr]&$10^7$ & $10^7$ & $10^{10}$ & $10^{10}$\\
\tableline
\tableline
\end{tabular}
\end{center}
\end{table}












\begin{table*}
\begin{centering}
\begin{tabular}{|c|c|c|c|c|c|c|c|c|c|}
\hline
\multicolumn{10}{|c|}{TABLE 4: Fraction of evolved binary systems significantly affected by Kozai-Lidov and tidal friction evolution.} \tabularnewline
\hline
\multicolumn{1}{|c|}{\textbf{Cusp Models}} & \multicolumn{9}{c|}{\textbf{Time (Myrs)}}\tabularnewline
\hline 
Cusp - $\gamma=0.5$  & 5  & 10  & 50  & 100  & 500  & 1000  & 5000  & 10000  & 12000\tabularnewline
\hline 
\hline 
Stellar type change 1  & 0 & 0.01 & 0.03 & 0.04 & 0.07 & 0.08 & 0.10 & 0.11 & 0.11\tabularnewline
\hline 
Stellar type change 2  & 0 & 0.01 & 0.04 & 0.06 & 0.10 & 0.11 & 0.12 & 0.13 & 0.13\tabularnewline
\hline 
Period change  & 0.05 & 0.07 & 0.10 & 0.10 & 0.13 & 0.13 & 0.14 & 0.14 & 0.14\tabularnewline
\hline 
Mass change 1  & 0 & 0 & 0.01 & 0.03 & 0.06 & 0.09 & 0.23 & 0.27 & 0.28\tabularnewline
\hline 
Mass change 2  & 0 & 0.02 & 0.12 & 0.17 & 0.28 & 0.34 & 0.48 & 0.50 & 0.51\tabularnewline
\hline 
Total evolutionary changes  & 0.05 & 0.08 & 0.20 & 0.26 & 0.39 & 0.47 & 0.66 & 0.71 & 0.72\tabularnewline
\hline 
\end{tabular}%

\begin{tabular}{|c|c|c|c|c|c|c|c|c|c|}
\hline 
Cusp - $\gamma=1.5$  & \multicolumn{9}{c|}{}\tabularnewline
\hline 
\hline 
Stellar type change 1  & 0 & 0.01 & 0.04 & 0.06 & 0.10 & 0.10 & 0.12 & 0.13 & 0.13\tabularnewline
\hline 
Stellar type change 2  & 0 & 0.02 & 0.04 & 0.07 & 0.11 & 0.13 & 0.14 & 0.15 & 0.15\tabularnewline
\hline 
Period change  & 0.06 & 0.07 & 0.10 & 0.13 & 0.15 & 0.15 & 0.16 & 0.17 & 0.18\tabularnewline
\hline 
Mass change 1  & 0 & 0 & 0.02 & 0.03 & 0.10 & 0.15 & 0.28 & 0.31 & 0.32\tabularnewline
\hline 
Mass change 2  & 0 & 0.03 & 0.15 & 0.22 & 0.38 & 0.46 & 0.57 & 0.58 & 0.58\tabularnewline
\hline 
Total evolutionary changes  & 0.06 & 0.09 & 0.24 & 0.32 & 0.52 & 0.61 & 0.77 & 0.80 & 0.80\tabularnewline
\hline 
\end{tabular}
\par\end{centering}

\begin{centering}
\begin{tabular}{|c|c|c|c|c|c|c|c|c|c|}
\hline 
\textbf{Disk Models}  & \multicolumn{9}{c|}{\textbf{Time (Myrs)}}\tabularnewline
\hline 
Disk - $\gamma=0.5$  & 5  & 10  & 50  & 100  & 500  & 1000  & 5000  & 10000  & 12000\tabularnewline
\hline 
Stellar type change 1  & 0.01 & 0.08 & 0.08 & 0.07 & 0.07 & 0.07 & 0.08 & 0.08 & 0.08\tabularnewline
\hline 
Stellar type change 2  & 0.04 & 0.14 & 0.14 & 0.13 & 0.13 & 0.13 & 0.14 & 0.14 & 0.14\tabularnewline
\hline 
Period change  & 0.05 & 0.18 & 0.20 & 0.20 & 0.20 & 0.20 & 0.21 & 0.21 & 0.21\tabularnewline
\hline 
Mass change 1  & 0 & 0.01 & 0.01 & 0.02 & 0.02 & 0.02 & 0.04 & 0.04 & 0.04\tabularnewline
\hline 
Mass change 2  & 0.03 & 0.23 & 0.26 & 0.26 & 0.26 & 0.26 & 0.26 & 0.26 & 0.26\tabularnewline
\hline 
Total evolutionary changes  & 0.06 & 0.33 & 0.38 & 0.38 & 0.38 & 0.38 & 0.40 & 0.40 & 0.40\tabularnewline
\hline 
\end{tabular}
\par\end{centering}

\begin{centering}
\begin{tabular}{|c|c|c|c|c|c|c|c|c|c|}
\hline 
Cusp - $\gamma=1.5$  & \multicolumn{9}{c|}{}\tabularnewline
\hline 
\hline 
Stellar type change 1  & 0.01 & 0.08 & 0.15 & 0.15 & 0.15 & 0.13 & 0.15 & 0.16 & 0.16\tabularnewline
\hline 
Stellar type change 2  & 0.04 & 0.15 & 0.19 & 0.19 & 0.19 & 0.18 & 0.19 & 0.20 & 0.20\tabularnewline
\hline 
Period change  & 0.06 & 0.18 & 0.19 & 0.19 & 0.19 & 0.18 & 0.19 & 0.19 & 0.19\tabularnewline
\hline 
Mass change 1  & 0 & 0 & 0.08 & 0.08 & 0.08 & 0.08 & 0.10 & 0.11 & 0.11\tabularnewline
\hline 
Mass change 2  & 0.03& 0.26 & 0.27 & 0.27 & 0.27 & 0.27 & 0.27 & 0.27 & 0.27\tabularnewline
\hline 
Total evolutionary changes  & 0.07 & 0.37 & 0.41 & 0.41 & 0.41 & 0.41 & 0.44 & 0.44 & 0.44\tabularnewline
\hline 
\end{tabular}
\par\end{centering}


%
%
\end{table*}


\subsection{Post KCTF Binary stellar evolution}\label{sub:post-MS}

Once the binaries go off the main sequence and become giants, their
radius will significantly expand. 
The high eccentricities induced via the KL mechanism, combined with  such an expansion in the size of each inner binary star, could lead to efficient production of coalescing/strongly-interacting post-MS binaries. Accounting
for the effects of the KCTF on the long term stellar evolution of the binaries
is therefore an important step.

In Table 4  we show the fraction of binary systems in
which stellar evolution was significantly affected by their earlier
KCTF evolution. We evolve 1304 binaries with $\gamma=0.5$ and 846 binaries with $\gamma=1.5$ originating from the cusp; and 106 binaries with $\gamma=0.5$ and 89 binaries with $\gamma=1.5$ originating from the disk. In the cusp models the stellar evolution
of the majority of the binaries is significantly
affected; rising from 45-60 \% after one Gyr and up to 70-80 \% after
10 Gyrs of evolution. In the disk models KCTF significantly affected a large
fraction (30-40 \%) of the binaries already during the first 10 Myrs of evolution, i.e. it should have affected the observed stellar population in the GC stellar-disk which is of comparable age. Also shown is the the later evolution of stellar-disk binaries beyond 10 Myrs (which could be relevant
to past epochs of disk star formation), though most of the effects 
already arise by 10 Myrs.

The time $t=0$ in table 4 corresponds to the beginning of the of the evolutionary MS evolution (i.e. all binary components begin as ZAMS stars). However, note that the KCTF binaries begin their evolution with the orbital parameters AFTER the KCTF evolution. In other words their evolutionary time is reset to zero, while the orbital parameters are taken after the KCTF evolution on the MS. This inconsistency does not affect the results significantly, as the main KCTF evolution occur on time scales which are (typically) much shorter than the evolutionary timescales; in particular letting the binaries evolve with the additional MS time, but with their new post-KCTF orbits make little affect on their evolution compared with their post-MS evolution.

 Note that in some cases the fractions of KCTF-affected binaries slightly drop at some evolutionary time compared to earlier times. This can arise, for example, when a secondary component in a post-KCTF binary evolves faster than the corresponding star in the binary evolved in isolation (non-KCTF). Such a star may change its stellar type, e.g. it accreted mass, became more massive and evolved faster to become a WD at some stage. The same component in this example may also evolve later-on to become a WD. For this reason the systems may be counted as differing (in terms of the stellar type of the secondary component) at an early stage when the star is a WD in the post-KCTF case, but its corresponding non-KCTF evolved star is still a red-giant. At the later stage, when the this non-KCTF case also evolves to a WD, the stars are counted as similar in terms of their stellar type, but may be counted as differing in their masses. Therefore, once we account for all possible changes included (stellar type, mass and period)\footnote{a post-KCTF system is counted as differing from its corresponding non-KCTF one if at least one of these elements change, and is not counted twice if more than one change is observed; hence the ``total evolutionary changes'' row in Table 4, is not a simple sum of the rows above},  the overall fraction of affected KCTF-binaries is a monotonically rising function, and the analysis captures the overall magnitude of changes due to the early KCTF evolution.

\subsection{N-body integration}\label{nbody}


\begin{table*}
\begin{centering}
\begin{tabular}{l|l|l}
\tableline
\multicolumn{3}{c}{TABLE 5. Inner binary parameters in N-body runs} \\
\tableline
\tableline
Symbol & Definition & Distribution\\ \hline
$m_{1,2}$ & stellar mass & IMF with $\alpha=1.7$\\
$a_1$ & Inner binary semimajor axis & lognormal with  $<logP(d)>=4.8$  and $\sigma(d)=2.8$\\
$e_{1, 0}$ & Inner binary initial eccentricity & thermal\\
$i_{init}$ & Initial mutual inclination & uniform in ($75^o$, $105^o$) \\
$R_{1,2}$ & stellar radius & $R_{1,2}=(m_{1,2}/M_{\bigodot})^{0.75}R_{\bigodot}$\\
$r_{p,out}$, Set 1& external orbit periapse& $5\times r_{bt}$\\
$r_{p,out}$, Set 2 & external orbit periapse& uniform in $(0, 10\times r_{bt}$)\\
$r_{apo, out}$ & external orbit apoapse & 0.03 pc\\ 
$k_2$ & Tidal Love number & $0.028$\\
$Q$ & Tidal dissipation factor & $10^6$\\
\tableline
\tableline
\end{tabular}
\par
\end{centering}
\end{table*}



The orbit average approximation used in \ref{sec:ktcf} is based on the assumption that the
binary angular momentum, $\ell_{1}=\sqrt{1-e_1^2}$, changes on timescales that are
longer than both inner and outer binary orbital periods. As discussed below,
the fact that $\ell_{1}$ can change on a timescale short compared
to the inner binary period has special implications for our study. The condition
that the binary angular momentum changes by order of itself between
two consecutive periapsis passages can be expressed in terms of the
system SMAs and eccentricities as \citep{2014ApJ...781...45A}:
\be
\sqrt{1-e_{1}}\lesssim5\pi\frac{M_{\bullet}}{M_{b}}\left[\frac{a_{1}}{a_{out}(1-e_{out})}\right]^{3}.\label{eq:sp_om2}
\ee

If the binary eccentricity satisfies this condition than the orbit can evolve and reach $e_{1}\sim1$, i.e. a colliding
trajectory, before post--Newtonian (PN) and tidal terms can limit the maximum eccentricity
attainable during a KL cycle. Thus, we expect some systems that
do not merge when using the double averaged approach, to merge when
they are evolved through direct integration of the equations of motion.

We note that oscillations occurring on the orbital timescale of the
outer orbit can be also important to the evolution~\citep{2014MNRAS.439.1079A}.
However, the condition given by equation~(\ref{eq:sp_om2}) turns out to be the
most relevant for our study given that the majority of merging systems
can collide only if they experience a clean collision, i.e., a collision
where dissipative and non dissipative tidal and (PN) terms do not play
an important role. More specifically, the condition for such an event 
requires that changes in the binary angular momentum occurring over the
orbital period of the binary are of the same order of, or larger than the angular
momentum associated with the scale at which other dynamical processes~(e.g.,
tidal dissipation or GR precession) can affect the evolution.
For example, considering a conservative dissipation scale $\tilde{r}=2(R_{1}+R_{2})$,
we find~\citep{2014ApJ...781...45A}: 
\be
\frac{r_{p,out}}{r_{bt}}\lesssim10\times\left(\frac{a_1}{{\rm AU}}\frac{10^{11}{\rm cm}}{\tilde{r}}\right)^{1/6}~.\label{eq:c-diss}
\ee
In this region of parameter space, where $\ell_{1}$ can change on the timescale of order of binary period, the orbit average approximation
cannot accurately describe the binary dynamics. 

In order to determine the likelihood for a stellar merger within the
region of parameter space where the orbit average technique is less
accurate, we carried out a number of direct integration of three-body
systems. These consist of binaries with orbital periapsis distance
to the MBH $a_{out}(1-e_{out})/\lesssim10r_{bt}$. It is from these
systems, which experience a relatively close approach to the MBH,
that we expect the largest number of mergers, as well as a larger
discrepancy between the results of the three-body integration and
the predictions of the orbit average equations~(e.g., equation~\ref{eq:c-diss}).

The triple systems were evolved using the high accuracy integrator
AR-CHAIN \citep{2008AJ....135.2398M} which includes PN corrections up
to order 2.5; to these we added terms corresponding to apsidal precession
due to tidal bulges and tidal dissipation. We model tidal effects
using the formulation given in equations (12) and (13) of \citet{1998MNRAS.300..292K}.
Approximately 1000 systems were evolved for a time of $10\times T_{Kozai}$,
and considered as mergers if during the evolution the stars approached
each other within a distance $r \leq R_{1}+R_{2}$, i.e., their separation
became smaller than the sum of their radii.

The parameters describing the mass and orbital distributions from
which the initial conditions of the 3-body runs were drawn are given
in Table 5. We adopted a fixed orbital apoapse of $0.03~$\pc, which
corresponds approximately to the inner radial extent of the stellar
disk at the Galactic center. Given its small apoapse, the eccentricity of
the external orbit was chosen such that the periapsis distance to
the MBH was uniform within the radial range $r_{p,out}\leq10r_{bt}$,
or such that $r_{p,out}=5r_{bt}$.

The results of the simulations are shown in Table~6. For the highly
inclined systems considered, the likelihood of a merger is $\approx50\%$.
Accounting for all possible inclinations, and for $r_{p,out}\leq10r_{bt}$
the merger probability is $\approx0.16$. We note that our results
are similar to those obtained by \citet{2010ApJ...713...90A}. Our initial conditions
and integrator are essentially the same as theirs with the difference
that the equations of motion of \citet{2010ApJ...713...90A} did not include terms
accounting for dynamical effects due to tides. The similarity between
the results of the two papers suggests that tides (as well
as PN terms) have essentially no effect on the binary dynamics as
expected if the binary angular momentum evolves substantially on a
timescale of the order the binary orbital period.

\begin{figure}
\centering
\includegraphics[angle=0,width=2.9in]{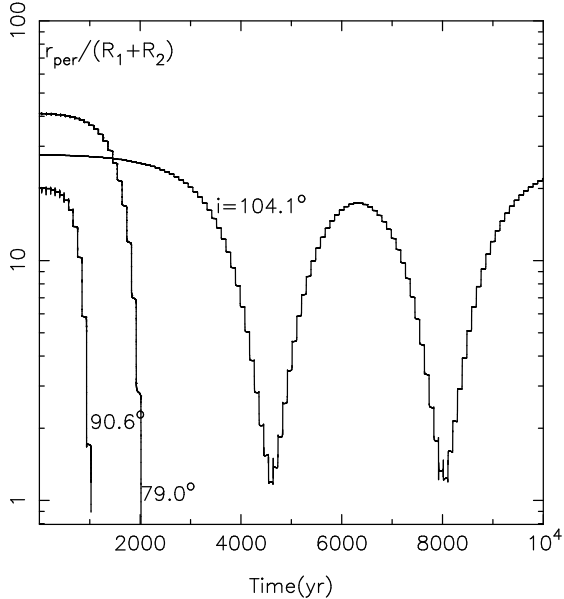}
\caption{Example of  N-body runs. The periapsis separation of the binary to the MBH was set to $r_{p,out}=5r_{bt}$.
 When $r_{p,in}/(R_1+R_2) < 1$ systems are considered to have merged. $i_0=90^{o}.6$ corresponds to $m_1=1M_{\odot}$, $m_2=0.8M_{\odot}$, $a_{1}=0.38AU$; $i_0=79^{o}$ corresponds to $m_1=3M_{\odot}$, $m_2=3M_{\odot}$, $a_{1}=0.45AU$; $i_0=104^{o}.1$ corresponds to $m_1=1M_{\odot}$, $m_2=0.6M_{\odot}$, $a_{1}=0.26AU$. 
}  \label{Fig:Nbody_e}
\end{figure}

A few examples are given in figure \ref{Fig:Nbody_e}. The binary eccentricity increases
as it orbits the Galactic center. Given the large eccentricity of
the external orbit most of the evolution occurs during the binary
closest approach to the MBH where the gravitational interaction is
the strongest. The binary angular momentum receives a \textquotedbl{}kick\textquotedbl{}
at each periapse passage, with two consecutive jumps separated roughly
by the external orbital period of the binary. The step size of the
angular momentum kick increases roughly as $\Delta\ell\sim\sqrt{1-e_{1}}$.
These systems experience a clean ``head-on'' collision and their dynamics
can be  model quite accurately  as it was purely Newtonian.


\begin{table}[h]
\begin{center}
\begin{tabular}{l|l|l}
\tableline
\multicolumn{3}{c}{TABLE 6. Results of N-body simulations} \\
\tableline
\tableline
fraction of systems & $r_{p,out}=5r_{bt}$ &$r_{p,out}\leq10r_{bt}$ \\ \hline
 ''mergers" & $0.61$ & $0.54$\\
disrupted  & $0$ & $0.12$ \\
 binaries & $0.39$ & $0.46$\\
\tableline
\tableline
integration $T_{max}$ &$10\times T_{Kozai}$ & $10\times T_{Kozai}$\\
\tableline
\tableline
\end{tabular}
\end{center}
\end{table}



\section{DISCUSSION}\label{sec:discussion}

Our results show that binary secular evolution could play a major
role in their stelar evolution both on the MS as well as during their post-MS
lifetime. In the following we briefly discuss several possible implications.

\subsection{Merger products}\label{sub:MS-merger}

\subsubsection{Blue stragglers and the observed mass function in the GC}

As shown in section \ref{sec:numerics}, KCTF evolution and quasi-secular
evolution can lead to binary mergers and collisions on the main-sequence.
Such merger/collision products become
``rejuvenated'' stars, more massive than each of their original
progenitors, and could possibly be observed as blue stragglers, as was
originally suggested in \citep{2009ApJ...697.1048P, 2009ApJ...698.1330P} and further investigated by \citet{2011ApJ...731..128A} through hydrodynamical simulations..
However, given the complex stellar population of the GC showing evidence
for continuous star formation, detecting such stars as blue-stragglers
could be challenging. We note that such evolution will also lead to
the formation of young massive merger products and might affect the
observed mass-function of the OB stellar populations and further bias
it to be top-heavy, and possibly explain the origin of the most massive S-stars.

\subsubsection{Post main-sequence evolution}

As discussed in Section \ref{sub:post-MS} and shown in Table
4, KCTF evolution significantly affects the post-MS evolution
of most (a significant fraction) of the KCTF-evolved binaries in the
cusp (disk) models, leading to strong interactions between the binary
components, mostly through mass-transfer and/or common envelope evolution.
The exact actual outcome of such evolution depends on the secular
dynamics that can still affect the binaries during their post-MS stage,
which is not modelled here. However, these results already clearly show
that the fraction of strongly-interacting binaries is likely
to be significantly enhanced among stars in the GC close to the MBH.
Such strongly interacting binaries could manifest themselves through a wide variety of outcomes,
from progenitors of type Ia SNe, mass-transfer rejuvenated stars (i.e.
in addition to the merger-produced rejuvenated stars in Section \ref{sub:MS-merger}),
as well as gravitational-wave sources (discussed in \citet{2012ApJ...757...27A})
and X-ray sources in accreting binaries, and may help explain the
observed overabundance of the latter \citep{mun+05}.

\subsection{G2-like objects as stellar merger products}

\citet{2012Natur.481...51G, 2013ApJ...763...78G} reported the discovery of G2, an extremely red object initially interpreted by these authors as a ~ 3 Earth-mass gas cloud on a highly eccentric orbit with a closest approach to our Galaxy's central black hole expected to occur in mid March 2014. The orbit of G2 has an eccentricity of 0.98 and SMA of ~ 0.03 pc.  If the gas cloud interpretation is correct, G2 should be turned apart by the tidal field of the MBH as it passes through periapsis at a distance of 130 AU, potentially allowing us to observe with an unprecedented level of details an accretion event onto a MBH. 

A number of authors have challenged the interpretation of G2 as a gas cloud and proposed instead alternative scenarios which invoke an underlying star ~\citep{2012ApJ...756...86M, 2012NatCo...3E1049M, 2013ApJ...768..108S}. A stellar nature of G2 also appears to be in agreement with recent observations obtained on 2014 March 19 \& 20 (UT). These observations show that G2, currently experiencing its closest approach, is still intact, in contrast to predictions for a simple gas cloud hypothesis (Ghez et al. 2014). \citet{2013ApJ...773L..13P} proposed that although G2 does have a gaseous component that is tidally interacting with the central black hole, there is likely a central star providing the self-gravity necessary to sustain the compact nature of the object. These authors argued that the  G2's observed physical properties (red, compact, and at times marginally resolved) are  consistent with the expected observables for stars which have recently undergone a merger. 
Following   \citet{2013ApJ...773L..13P}  we investigate here the possibility that G2 is the result of the merger of two stars. In our scenario, the G2 progenitors were initially part of a binary  that  was orbited by a distant third object, i.e., the initial system was a triple at an initial large distance from the GC. The triple was then scattered  onto a quasi-radial orbit towards the GC through gravitational interactions with other stars or
massive perturbers~\citep{2009ApJ...698.1330P}. At the closest approach with the MBH the triple was dissociated leaving a binary onto 
an extremely eccentric  and inclined orbit around the MBH. Such a binary merged
due to the KL oscillations induced by its gravitational interaction with the MBH, producing
an object with the peculiar features of G2. 

A triple system that enters its tidal disruption radius leaves in $\sim 50\%$ of all cases a binary star around the MBH -- in the other $50\%$ of the cases the binary is ejected and its companion is captured~\citep{2006ApJ...653.1194B}. 
The initial SMA  of the triple, $a_{tr}$, is obtained by requiring  that the
binary is left onto an orbit with external orbital periapsis roughly  
equal  the tidal disruption radius of the triple, $r_{d,tr}=r_{p, out}\approx 130~$AU; this gives
\begin{equation}
a_{tr}\sim 1 \left( \frac{M_{tr}}{M_{\bullet}} \frac{4\times 10^6 M_{\odot}} {3M_{\odot}}\right)^{1/3} {\rm AU}~.
\end{equation}
with $M_{tr}$ the total mass of the triple system.
 By requiring the triple to be initially stable~\citep{2001MNRAS.321..398M}, we obtain an  upper  limit to the 
 semi-major axis  of the binary $a_1$:
 \begin{eqnarray}
a_{1} &\lesssim & \frac{a_{tr}~(1-e_{tr})}{3.3} \left[ \frac{2}{3}\left(1+\frac{m_{3}}{M_{b}}\right) 
\frac{1+e_{tr}}{\left(1-e_{tr}\right)^{1/2}}\right]^{-2/5} \\
&=& 0.3  \frac{a_{tr}~(1-e_{tr})}{ {\rm AU}} \left[ \frac{2}{3}\left(1+\frac{m_{3}}{M_{b}}\right) 
\frac{1+e_{tr}}{\left(1-e_{tr}\right)^{1/2}}\right]^{-2/5}  {\rm AU}~,  \nonumber \\ \nonumber
\end{eqnarray}
with $e_{tr}$ the eccentricity of the outer companion with respect to the inner binary center of mass, $M_b$ total mass of the inner binary and $m_3$ the mass of the third (ejected) star.
The triple star parameters -- an outer orbital distance of a few AU and a fraction of AU inner binary separation -- 
required to give the observed orbit of G2 are therefore quite reasonable.
For example if we assume a constant probability per $\ln(a_1)$ for $0.02<a_1<20~$AU
 then the probability of finding a binary in the range of $0.02-0.1$AU is $10\%$; 
\citet{1981ApJ...246..879F} finds that a fraction of 0.2 of the
more massive systems in close multiple stars have outer binary periods shorter than half a
year, corresponding to $a_{tr} <1$AU.

The binary will be left onto an orbit with a periapsis distance from the MBH that is  a few times its
tidal disruption radius.  The SMA of the captured binary is:
 \begin{equation}
a_{G2} \simeq  0.017 q \left( \frac{M_{\bullet}}{4\times 10^6 M_{\odot}}\right) 
\left( \frac{1000 {\rm km~s^{-1}}}{v_{ej}} \right)^2 {\rm pc}~,
 \end{equation}
 where $q=M_{G2}/m_{3}$, with $M_{G2}$ being a mass of G2 and corresponding to $M_b$, and $v_{ej}$ is the velocity at infinity of the ejected star.
In agreement with a triple disruption origin for G2,
its orbit SMA is comparable to the orbital radii inside of the S-stars that are thought
 to be  deposited at the GC  through binary disruptions by the MBH~\citep{1988Natur.331..687H}.
 The characteristic eccentricity of the binary orbit  depends on $q$ and the
  mass ratio of the binary to the MBH~\citep{2013ApJ...768..153Z}:
   \begin{equation}
e_{G2}\sim 1-\frac{2.8}{q^{1/3}(1+q)^{2/3}}\left(\frac{M_{G2}}{M_{\bullet}}\right)^{1/3},
    \end{equation}
    which gives $e_{G2}=0.98$ for $q=1$ and $M_{G2}/M_{\bullet}=5\times 10^{-7}$.
  It is notable that the   
    triple disruption scenario followed by a merger of  the captured binary,  naturally 
reproduces G2 observationally derived SMA, $\sim0.03$~pc, and eccentricity, $0.98$.
The orbital evolution of the binary following its capture around the 
MBH will resemble that of the systems of Fig. ~\ref{Fig:Nbody_e} -- the two stars will
merge on a timescale of order $\sim10^3-10^4~$yr, provided that the mutual inclination between outer and inner orbit is large.

We add that the observed properties of G2 are not far different from those of 
several observed stellar objects that are thought to have formed through a stellar merger.
For example, G2 temperature and radius are similar to those of
BLG-360 recently observed by \citet{2013A&A...555A..16T}. Also V4332 Sgr~\citep{2010A&A...522A..75K} as
well as V1309 Sco~\citep{2013ApJ...777...23Z} at present are strong IR sources with a similar dust temperature as G2. 
Also the well known red-nova V838 Mon~\citep{2009ApJS..182...33K} 
although looks different from G2, as an M6 component dominates,  it is
also bright in IR and it is quite probable that if oriented in a different
way than observed, i.e. that denser dust region obscures the central
object, it would also be observed  as similar to G2. CK Vul (Nova Vul 1670)
it is also likely to have been a stellar merger and now is seen only as an
infrared source~\citep{2003A&A...399..695K}. If correctly interpreted, CK Vul  shows that the
evolution of stellar merger remnants are slow, so
the G2 merger could have happened hundreds or thousands of years ago. In
fact, when two similar stars merge the remnant resembles a
pre-MS star and evolves on a similar time scale. 

The main question is whether these mergers are common enough such that there is a finite
probability of observing one object with the characteristics of G2 at any time in the GC.
We evaluate the event rate for mergers produced by triple disruptions followed by KL evolution of the
captured binary as:
\begin{equation}
\Gamma_m=\Gamma_b \times f_{tr}  \times  f_{b}  \times f_{m}~,
\end{equation}
where $\Gamma_b$ is the tidal disruption event rate for binaries at the GC, 
 $f_{tr}$ is the fraction of binaries in triple systems, $f_{b}$ is the fraction of events that leave a binary 
 around the MBH and $f_{m}$ is the fraction of these systems that end up merging.
We take $\Gamma_b=10^{-4}{\rm yr}^{-1}$ as this is roughly the production rate required to obtain 
 the observed number of S-stars ($\sim 20$) within a radius of $0.05$~pc of SgrA*~\citep{2007ApJ...656..709P}.
 From the simulations of Section~\ref{nbody} we have $f_{m}\approx 0.1$. We set $f_b=0.5$ and
 following \citet{2009ApJ...698.1330P} we assume a triple fraction of $f_{tr}=0.2$.  With these values we find 
 $\Gamma_m\approx 10^{-6}{\rm yr}^{-1}$. The merger product 
 can look like G2 only before it contracts back to the MS and continues its evolution
 as a normal MS star. The thermal timescale over which  the star will reach the main-sequence
is of order $\sim 10^{5}{\rm yr}$~\citep[see Table 6 of~][]{2011ApJ...731..128A}, so
the probability of finding a merger product out of thermal equilibrium 
at any time in the GC is roughly $\sim 0.1$. 
We note however that the continuous tidal interaction of the puffed-up merger product with the MBH could keep the star from  reaching thermal equilibrium. In such a scenario the merger product  could maintain properties similar to G2 for a time much longer than $10^{5}{\rm yr}$, increasing its chance of being observed with such properties.

\section{SUMMARY }

\label{sec:conclusions}

In this study we explored the secular evolution of main sequence
binaries around a massive black hole taking into account tidal effects
of the inner binary and the interaction of the binaries with the stellar
environment. The later stellar evolution of the dynamically evolved
binaries was also explored, albeit simplistically. We considered binaries
in the observed young stellar disk, as well as binaries in the old
stellar cusp. For both cases we took into account two possible density
profiles; a cusp,  and a core-like profile. We find that the MBH can
induce very high eccentricities in the orbits of binaries around it, causing
them to coalesce or significantly evolve through tidal evolution and/or
later stellar evolution (common envelope, mass transfer etc.), on timescales
shorter than the evaporation time of the binaries around the MBH. The
main results of our study are summarized below:
\begin{itemize}
\item The dynamics of binaries with high inclinations with respect
to their orbit around the MBH ($70^{\circ}\lesssim i\lesssim110^{\circ}$)
are strongly affected by KCTF leading to the shrinking of the binary SMA,
 at distances up to 0.5 pc from the MBH. As a consequence of a shorter SMA,
the binaries' evaporation time becomes  longer and their 
 survival rate  is enhanced by $\sim 10\%$ in the GC environment.
\item 2-3 \% of all binaries are likely to merge due to secular
evolution on the main sequence. Mergers during the main-sequence evolution
of the binaries typically occur through high excitation of eccentricity
leading to direct physical collision before tidal dissipation can become important to the evolution. 
\item The post-MS evolution of the majority of the binaries that suffered secular evolution during the MS phase is significantly altered by KCTF which can lead
to strong binary interaction through mass-exchange and/or common-envelope
evolution. As a result, this  could
enhance the fraction of X-ray sources in the GC, and may offer an explanation for
the overabundance of X-ray sources in the GC. 
\item Merger products and mass-transfer systems could produce rejuvenated
stars (``blue stragglers'') and affect the observed mass function
of the GC stars in both the young disk and the stellar cusp.
\item We suggest that the recently observed G2-cloud in the GC could
potentially be the product of a binary merger due to KL cycles induced by the central MBH.
\end{itemize}

\acknowledgments Authors are grateful to Cole Miller, Romuald Tylenda and Noam Soker for helpful discussions.This research has made use of the SIMBAD database, operated at CDS,
Strasbourg, France, and of NASA's Astrophysics Data System. SP is supported in part by NSERC of Canada. HBP acknowledges support from the Technion Deloro fellowship and the I-CORE Program of the Planning and Budgeting Committee and The Israel Science Foundation grant 1829/12.

\bibliography{1820}{}

\end{document}